\documentclass{ws-ijmpd}

\usepackage{graphicx}
\usepackage{amssymb}

\begin{document}

\markboth{T. Padmanabhan}
{Gravity as elasticity of spacetime}

%
\catchline{}{}{}{}{}
%

\title{Gravity as elasticity of spacetime: a paradigm to understand\\
horizon thermodynamics and cosmological constant\footnote{This essay received an ``honorable mention'' 
in the 2004 Essay Competition of the Gravity Research Foundation -- Ed.}}

\author{T. Padmanabhan}

\address{IUCAA,
Post Bag 4, Ganeshkhind, \\
Pune - 411 007, INDIA.\\
E-mail address: nabhan@iucaa.ernet.in}

\maketitle

\begin{history}
\received{Day Month Year}
\revised{Day Month Year}
\end{history}

\begin{abstract}

It is very likely that the quantum description of spacetime is quite different from  what we
perceive at large scales, $l\gg (G\hbar/c^3)^{1/2}$. The long wavelength description of spacetime, based on Einstein's equations, is similar to the description of a continuum solid made of a large number of microscopic degrees of freedom. This paradigm provides a novel interpretation of coordinate transformations as
deformations of ``spacetime solid" and allows one to obtain Einstein's equations as a consistency condition in the long wavelength limit.  The entropy contributed by the microscopic degrees of freedom reduces to a pure surface contribution when Einstein's equations are satisfied.
The horizons arises as ``defects" in the ``spacetime solid" (in the sense of well defined singular points) 
and contributes an entropy which is one quarter of the horizon area.
Finally, the response of the microstructure  to vacuum energy leads to a near cancellation of the cosmological constant, leaving behind a tiny fluctuation which matches with the observed value.

\end{abstract}


\noindent

The quantum description of spacetime is likely  to be as different from the classical description, as the atomic description of a solid is from the macroscopic continuum description of a solid.  The latter uses concepts like density, stress and strain, bulk flow velocity etc., none of which has much relevance in the microscopic scale; the quantum description 
of molecules in a solid cannot be obtained by quantising the classical macroscopic variables. The entire
language (and the  variables) need to be different, with the continuum description arising in
a specific limit. Similarly, we will expect the variables and the language used in the quantum description of spacetime to be quite different from the description  of the spacetime continuum using, say,
the metric tensor.  Quantisation of the metric has as much relevance --- in this paradigm --- as quantising, say,  the density and bulk flow velocity of a solid  with the hope of  obtaining a quantum theory of molecules.

While the full microscopic theory cannot be obtained without further inputs, this paradigm (``continuum spacetime is like an elastic solid"), {\it taken seriously},
brings about a new and powerful perspective regarding the nature of the spacetime. As we shall show 
{\it this perspective allows one to obtain the Einstein's equations in a very novel manner and throws light on several issues in semiclassical gravity like the thermodynamics of horizons and the cosmological constant.}  

Let us begin by noting that, in
      the study of elastic deformation in continuum mechanics \cite{ll7}, one begins
      with the deformation field $u^\alpha (x) \equiv \bar x^\alpha -x^\alpha$ which indicates how 
      each point in a solid moves under a deformation.
      (The Greek letters go over the spatial coordinates 1-3 while the Latin letters go over 0-3.)
      The deformation contributes to the thermodynamic functionals like free energy, entropy etc. In the absence
of  external fields, a constant $u^\alpha$ cannot make a contribution because of translational invariance. Hence,
to the lowest order, 
the thermodynamical functionals  will be quadratic in the scalars constructed from
    the derivatives of the deformation field.
       The derivative $\partial_\mu u_\nu$ can be decomposed into
      an anti symmetric part, symmetric traceless part and the trace corresponding to deformations
      which are rotations, shear and expansion. Since the overall rotation of the solid will  not  
     change the thermodynamical variables, only  the other two components, $S_{\mu\nu}
      \equiv \partial_\mu u_\nu +
      \partial_\nu u_\mu-(1/3)\delta_{\mu\nu}\partial_\alpha u^\alpha$ and $\partial_\alpha u^\alpha$,
      contribute.
      The extremisation of the relevant functional (entropy, free energy ....) 
 allows one to determine the equations which govern
      the elastic deformations. 
      
      We take the point of view that the classical spacetime is the coarse grained limit of some ---
      as yet  unknown --- microscopic substructure which could  even be discrete in nature.
      These degrees of freedom are usually excited only at Planck energy scales and hence
      do not directly contribute to the bulk physics of the spacetime. (This is similar to the fact
      that degrees of freedom residing inside the atom or nuclei do not contribute to the 
      specific heat of solids  since they are unexcited at normal temperatures. )
      The analogue  of elastic deformations in the case of spacetime manifold will be the 
      transformation $x^a \to \bar x^a = x^a+v^a(x)$.  
    Our paradigm requires us to take this transformation to be of fundamental importance rather than
      as ``mere coordinate relabelling''. In analogy with the elastic solid, we will attribute a
`thermodynamic'  functional --- which we shall take to be the entropy, for reasons which will be clearer
as we proceed  ---
       with a given  spacetime deformation. This will be a quadratic functional of $Q_{ab} \equiv
      \nabla_a v_b$ in the absence of matter. The presence of matter will, however,  break the translational
      invariance and hence there could be a contribution which is quadratic in $v_a$ as
      well. We may therefore take the form of the entropy functional to be  
      \begin{equation}
      S=\frac{1}{8\pi}\int d^4x\, \sqrt{-g}\, \left[
      M^{abcd}   \nabla_a v_b  \nabla_c v_d + N_{ab} v^av^b\right]
      \label{freeenergy}
      \end{equation}
      where the tensors, $M^{abcd}$ and $N_{ab}$ are yet to be determined.  They can depend on other
coarse grained macroscopic variables like the matter stress tensor $T_{ab}$, metric $g_{ab}$, other geometrical tensors etc. 
     
      Extremising $S$ with respect to the deformation field $v^a$ will lead to the equation
      \begin{equation}
      \nabla_a (M^{abcd}  \nabla_c)v_d = N^{bd} v_d
      \label{basic}
      \end{equation}
      In the case of elasticity, one would have used such  an equation to determine the deformation
      field $v^a(x)$. Further, one would have demanded some reasonable conditions on the deformation
      like, say, the mapping  should be nonsingular and uniquely invertible. 
       But the situation is quite different in the case of spacetime.  Here,
      in the  coarse grained limit  of continuum spacetime physics, one requires
      {\it any} deformation $v^a(x)$ to be allowed in the spacetime {\it provided
      the background spacetime satisfies Einstein's equations}. Hence,  if our ideas are correct,
      we should be able to choose $M^{abcd}$  and $N_{ab}$ in such a way that 
       equation (\ref{basic}) leads to Einstein's equation when we demand that it should hold for
       any $v^a(x)$. 
       
       Incredibly enough, this requirement is enough to uniquely determine
       the form of  $M^{abcd}$ and $N_{ab}$ to be: 
       \begin{equation}
       M^{abcd} = g^{ad}g^{bc} - g^{ab}g^{cd} ; \quad N_{ab} = 8\pi \left(T_{ab} - \frac{1}{2} g_{ab} T\right)
       \label{mndef}
       \end{equation}
       where $T_{ab}$ is the macroscopic stress-tensor of matter.
       In this case, the entropy functional becomes
        \begin{eqnarray}
      S&=&\frac{1}{8\pi}\int d^4x\, \sqrt{-g}\, \left[(\nabla_a v^b)(\nabla_b v^a) - (\nabla_b v^b)^2
       + N_{ab} v^av^b\right]\nonumber\\
       &=& \frac{1}{8\pi}\int d^4x\, \sqrt{-g}\, \left[ {\rm Tr}\ (Q^2) - ({\rm Tr}\ Q)^2 
       + 8\pi \left(T_{ab} - \frac{1}{2} g_{ab} T\right)v^a v^b\right]
       \label{tr}
      \end{eqnarray}
      where $Q_{ab}\equiv\nabla_av_b$.
      The variation with respect to $v^a$ leads to the Eq.(\ref{basic}) which, on using Eq.(\ref{mndef}),
      gives:
      \begin{equation}
      (\nabla_a\nabla_b - \nabla_b\nabla_a) v^a = 8\pi \left(T_{ab} - \frac{1}{2} g_{ab} T\right) v^a
      \end{equation}
      The left hand side is $R_{ab}v^a$ due to the standard identity for commuting the covariant
      derivatives. Hence the equation can hold for arbitrary $v^a$ only if
\begin{equation}
       R_{ab} =8\pi \left(T_{ab} - \frac{1}{2} g_{ab} T\right)
\label{albie}
\end{equation}
      which is the same as Einstein's equations.  This elegant result is worth examining in detail:

To begin with, note that we did {\it not} vary  the metric tensor to obtain Eq.(\ref{albie}). In this approach,
$g_{ab}$ and $T_{ab}$ are derived macroscopic quantities and are not fundamental variables. Einstein's equations arise as a consistency condition, reminiscent of the way it is derived in some string theory models due to the vanishing of beta function \cite{strings}. While the idea of spacetime being an ``elastic solid" has a long history (starting from \cite{sakharov}),
all the previous approaches obtain a low energy effective action in terms of $g_{ab}$s which are then varied to get Einstein's equations. Our approach is {\it very} different but is a simple consequence of taking our paradigm seriously.

Second, this result offers a  new perspective on  general coordinate transformations which are  treated 
as akin to deformations in solids.  General covariance now arises as a macroscopic symmetry in the long wavelength limit, when the spacetime satisfies the Einstein's equations. In this limit,  the deformation should not change
the thermodynamical functionals.
      This is indeed true; the expression for the entropy in Eq.(\ref{freeenergy}) 
      reduces to a four-divergence   when Einstein's equations
      are satisfied (``on shell") making $S$  a surface term: 
      \begin{equation}
      S = \frac{1}{8\pi}\int_{\cal V} d^4x\, \sqrt{-g}\, \nabla_i ( v^b \nabla_b v^i - v^i \nabla_b v^b)
      =\frac{1}{8\pi}\int_{\partial{\cal V}} d^3x\, \sqrt{h}\, n_i( v^b \nabla_b v^i - v^i \nabla_b v^b)
\label{onsur}
      \end{equation}
      {\it The entropy of a bulk region $\mathcal{V}$  of spacetime  resides in its boundary
      $\partial \mathcal{V}$ when Einstein's equations are satisfied.} In varying  Eq.(\ref{freeenergy})
      to obtain Eq.(\ref{basic}) we keep this surface contribution to be a constant. 

This result has an important consequence. If the spacetime has microscopic degrees of freedom,
then any bulk region will have an entropy and it has always been a surprise why the entropy scales as
the area rather than volume. Our analysis shows that, the semiclassical limit, when Einstein's equations hold
to the lowest order, {\it the entropy is contributed only by the boundary term
and the system is holographic.} 
(This idea has been  developed in more detail in a series of previous papers; see Ref.~\refcite{tppapers}.)

 This result can be put in a more familiar setting by noticing that, 
      in the case of spacetime, there is one kind of ``deformation" which is rather special --- 
      the inevitable translation forward in time: $t\to t+\epsilon$. More formally, one can consider this as arising from $x^a\to x^a+u^a$ where $u^a$ is the unit normal to a spacelike hypersurface (``many fingered time"). In this case, one can
      foliate the spacetime and introduce the the extrinsic curvature $K_{ab}$ built from the time like normal $u_a$ and the corresponding
      acceleration, 
      \begin{equation}
      K_{ab}= - \nabla_a u_b -u_au^j\nabla_ju_b = - \nabla_a u_b -a_b u_a.
      \end{equation}
       Since $(\nabla_a u_b)(\nabla^b u^a) = {\rm Tr}\
      (K^2), (\nabla_au^a)^2 = ({\rm Tr}\ K)^2$, our expression for entropy agrees with the
      kinetic part  in ADM Hamiltonian ${\rm Tr}\ (K^2) - ({\rm Tr}\ K)^2 $ when $v^a = u^a$.
      The surface term in Eq.(\ref{onsur}) now becomes:
      \begin{equation}
       S = -\frac{1}{8\pi}\int_{\cal V} d^4x\, \sqrt{-g}\, \nabla_i ( Ku^i+a^i)
      =-\frac{1}{8\pi}\int_{\partial{\cal V}} d^3x\, \sqrt{h}\, n_ia^i 
\label{oldres}
      \end{equation}
 where $a^i\equiv u^j\nabla_ju^i$ is the acceleration and the last result is valid on surfaces for which
 $u^in_i=0$, which is of relevance for horizons.    
 It is easy to show that, in this case, we will get an entropy that is
       proportional to the area of any horizon, if the horizon arises as a singular point in the
       deformation field. 
Near any static horizon, one can choose the metric to be
\begin{equation}
ds^2\approx-\kappa^2 N^2 dt^2 +dN^2 +\sigma_{AB}dx^Adx^B
\end{equation}
where the horizon is at $N=0$ and $\sigma_{AB}dx^Adx^B$ is the transverse metric with $A,B=2,3$. In the Euclidean version of the integral in Eq.({\ref{oldres}), restricting the range of time integration to $(0,2\pi/\kappa)$ due to periodicity of the metric and using $N(n_ia^i)\to -\kappa$ on the horizon ${\cal H}$, we find\cite{tppapers})
\begin{equation}
 S  
=-\frac{1}{8\pi}\int_0^{2\pi/\kappa}dt\int_{\cal H} d^2x\, \sqrt{\sigma}\,(N n_ia^i)=\frac{1}{4}({\rm Horizon\; Area})
\label{areaentropy}
      \end{equation}
While it agrees with known results the interpretation is quite different. The deformation field corresponding to
time evolution hits a singularity on the horizon, which is analogous to a topological defect in a solid. The entropy is the price we pay for this defect.  Our paradigm, therefore
can handle singular and nonsingular transformations in the same footing, with the former leading to the correct entropy for horizons. This provides a totally new view on the entropy of horizons. [This view also has deeper implications for causality in quantum gravity which is not explored here].

Finally, let us consider the implications of our result for the cosmological constant.     
       If we take the expression in (\ref{freeenergy}) but restrict ourselves to vectors of constant
      norm (that is, $v^a v_a = $ constant), then the vector field does not couple to the cosmological constant!
      This arises from the fact that, if $T_{ab}=\rho g_{ab}$ with a constant $\rho$, then $T_{ab} - \frac{1}{2} g_{ab} T=-T_{ab}$
      and the coupling term $N_{ab}v^av^b$ for matter is proportional to $v^2$, which is a constant. 
      In this case, one can show that the variation leads to the equation:
      \begin{equation}
      R_{ab} - \frac{1}{4} g_{ab} R=8\pi(T_{ab} - \frac{1}{4} g_{ab} T)
      \label{unimod}
      \end{equation}
      in which both sides are trace free. Bianchi identity can now be used to show that
      $\partial_a(R+8\pi T)=0$, requiring $(R+8\pi T)=$ constant. Thus  cosmological constant arises
      as an (undetermined) integration constant in such models \cite{tppr}, and could be interpreted as a
      Lagrange multiplier that maintains the condition $v^2=$ constant.  This suggests that the effect of vacuum energy density
      is to rescale the length of $v^a$.  The quantum
    micro structure of spacetime at Planck scale is capable of readjusting itself, soaking up any
    vacuum energy density which is introduced ---  like a sponge soaking up water. 
    
    This also provides an interesting reason why we observe a small but nonzero cosmological constant.
    Since the process suggested above is inherently quantum gravitational,
    it is subject to quantum fluctuations at Planck scales. 
     The cosmological constant we measure corresponds to this  small 
    residual fluctuation  (the ``wetness of the sponge") and will depend on the volume of the spacetime region that is probed.  
    It is small, in the sense that it has been reduced from $L_P^{-2}$ to $L_P^{-2}(L_PH_0)^2$, 
    (where $L_P$ is the Planck length and $H_0$ is the current Hubble constant)
    which indicates the fact that fluctuations --- when measured over a large volume --- is small
     compared to the bulk value.   An implementation of this suggestion was made in
     ref.\cite{cqglambda} where it was shown that the net effect can be described by
     a  `scalar field  potential' $V(\phi) = 
     - L_P^{-4} \ln \left(\phi/ \phi_0\right)$ in the semiclassical limit. It is obvious that
     the rescaling of such a scalar field by $\phi \to q \phi$ is equivalent to adding a cosmological
     constant with vacuum energy $-L_P^{-4} \ln q$. Alternatively, any vacuum energy
     can be reabsorbed by such a rescaling.   Identifying this scalar field with the length of the vector field
     $v^a v_a$ provides a concrete mechanism for the same.  It was also shown in ref. \cite{cqglambda} that
     there are inevitable residual fluctuations in the cosmological constant of the order of:
     \begin{equation}
\Delta\Lambda={8 \pi L_P^2\over \Delta{\cal V}}= \left({8\pi \over \alpha^2}\right){1\over\sqrt{{\cal V}}}\approx  {8\pi \over \alpha^2} H_0^2
\label{dellamb}
\end{equation}
where 
   $\alpha$ is a numerical constant. 
This will give $\Omega_\Lambda= (8\pi/3\alpha^2)$ which will --- for example --- lead to $\Omega_\Lambda =(2/3)$ if $\alpha = 2 \sqrt{\pi}$ which agrees with the observed value.

 \end{document}